\documentclass[epj]{webofc}%
\usepackage{amsmath}
\usepackage[varg]{txfonts}
\usepackage{amsfonts}
\usepackage{amssymb}
\usepackage{graphicx}%
\providecommand{\U}[1]{\protect\rule{.1in}{.1in}}
\woctitle{MESON2018 - the 15$^\textrm{th}$ International Workshop on Meson Production, Properties and Interaction}
\begin{document}

\title{Chiral anomaly and strange-nonstrange mixing}
\author{Francesco Giacosa\inst{1,2}\fnsep\thanks{fgiacosa@ujk.edu.pl} }

%\selectlanguage{english}

%insert email only for speaker/presenter

\institute{
Institute of Physics, Jan Kochanowski University,
ul. Swietokrzyska 15, Kielce, Poland
\and
Institute for Theoretical Physics, J.W. Goethe University,
Max-von-Laue-Str. 1, Frankfurt am Main, Germany}

\abstract{As a first step, a simple and pedagogical  recall of the $\eta$-$\eta'$  system is presented, in which
the role of the axial anomaly,
related to the heterochiral nature of the multiplet of (pseudo)scalar states, is underlined. As a consequence, $\eta$ is close to the octet and
$\eta'$  to the singlet configuration.
On the contrary, for vector and tensor states, which belong to homochiral multiplets, no anomalous
contribution to masses and mixing is present. Then, the isoscalar physical states are to a very good approximation nonstrange and strange, respectively.
Finally, for pseudotensor states, which are part of an heterochiral multiplet (just as pseudoscalar ones),
a sizable anomalous term is expected:
$\eta_2 (1645)$ roughly corresponds to the octet and $\eta_2 (1870)$ to the singlet.}
\maketitle

\section{Introduction}

The meson $\eta^{\prime}\equiv\eta^{\prime}(958)$ is special: its large mass
and its flavor content are strongly influenced by the so-called axial anomaly
\cite{thooft,feldmann,escribano} (the classical $U(1)_{A}$ symmetry of QCD is
broken by quantum fluctuations). Roughly speaking, $\eta^{\prime}$ corresponds
to a flavor singlet, while $\eta\equiv\eta(547)$ to the octet. In Sec. 1, we
recall some basic features of the $\eta$-$\eta^{\prime}$ and we connect them
to the \textit{heterochirality} \cite{anomaly,salamanca} of pseudoscalar
states and their chiral partners, the scalar states\textit{.}

A natural question is if the axial anomaly affects other mesons.
Interestingly, it turns out that the axial anomaly does \textit{not} affect
the vector states $\omega(782)$ and $\phi(1020)$ and the tensor states
$f_{2}(1270)$ and $f_{2}^{\prime}(1525)$ (see Sec. 3): $\omega(782)$ and
$f_{2}(1270)$ are (almost purely) nonstrange and $\phi(1020)$ and
$f_{2}^{\prime}(1525)$ strange. This fact can be nicely understood by the
\textit{homochirality }of the corresponding chiral multiplets, which involve
left- and right-handed currents\textit{. }For homochiral multiplets, no
anomalous mixing is realized \cite{anomaly}.

Are there other mesons for which the anomaly plays a role? This seems to be
the case of pseudotensor mesons (Sec. 4). As shown in the phenomenological
study of Ref. \cite{adrian}, the mesons $\eta_{2}(1645)$ and $\eta_{2}(1870)$
roughly correspond to octet and singlet states (the mixing angle is similar to
the one of $\eta$ and $\eta^{\prime}$).\ The pseudotensor mesons belong to a
heterochiral multiplet (just as pseudoscalar states), hence one can understand
why the axial anomaly is relevant.

\section{Pseudoscalar sector}

First, we review some features of the pseudoscalar sector. We consider the
strange-nonstrange basis $\eta_{N}=\sqrt{1/2}(\bar{u}u+\bar{d}d)$, $\eta
_{S}=\bar{s}s$ and the octet-singlet basis $\eta_{8}=\sqrt{1/6}(\bar{u}%
u+\bar{d}d-2\bar{s}s)$, $\eta_{0}=\sqrt{1/3}(\bar{u}u+\bar{d}d+\bar{s}s)$. The
physical fields $\eta\equiv\eta(547)$ and $\eta^{\prime}\equiv\eta^{\prime
}(958)$ are a mixture of $\eta_{N}$ and $\eta_{S}$ (and, similarly, of
$\eta_{0}$ and $\eta_{8}$), according to:
\begin{equation}%
\begin{pmatrix}
\eta\\
\eta^{\prime}%
\end{pmatrix}
=%
\begin{pmatrix}
\cos\theta_{P} & \sin\theta_{P}\\
-\sin\theta_{P} & \cos\theta_{P}%
\end{pmatrix}%
\begin{pmatrix}
\eta_{N}\\
\eta_{S}%
\end{pmatrix}
\text{ , }%
\begin{pmatrix}
\eta_{0}\\
\eta_{8}%
\end{pmatrix}
=%
\begin{pmatrix}
\sqrt{2/3} & \sqrt{1/3}\\
\sqrt{1/3} & -\sqrt{2/3}%
\end{pmatrix}%
\begin{pmatrix}
\eta_{N}\\
\eta_{S}%
\end{pmatrix}
\text{ .} \label{mixing}%
\end{equation}
The determination of $\theta_{P}$ is one important aspect of the problem.

We now introduce the Lagrangian terms for masses and mixing. The flavour
invariant term is simply given by:
\begin{equation}
\mathcal{L}_{P,U(3)}=-\frac{m_{P}^{2}}{2}\left(  \eta_{N}^{2}+\eta_{S}%
^{2}\right)  =-\frac{m_{P}^{2}}{2}\left(  \eta_{0}^{2}+\eta_{8}^{2}\right)
\text{ .}%
\end{equation}
After spontaneous symmetry breaking $m_{P}^{2}\propto(m_{u}+m_{d})/2$, see
e.g. Ref. \cite{dick}. If only $\mathcal{L}_{P,U(3)}$ is taken into account,
one could use $\eta_{0}$-$\eta_{8}$ or $\eta_{N}$-$\eta_{S}$ (the
octet-singlet choice is mathematically preferable). Next, the fact that the
$s$-quark is more massive than the quarks $u$ and $d$ is taken into account by
the Lagrangian
\begin{equation}
\mathcal{L}_{P,S}=-\frac{\delta_{P,S}}{2}\eta_{S}^{2}\text{ , }%
\end{equation}
with $\delta_{P,S}=2(m_{K}^{2}-m_{\pi}^{2})$ ($m_{K}$ and $m_{\pi}$ are the
kaon and pion masses). If $\mathcal{L}_{U(3)}+\mathcal{L}_{S}$ is considered,
the physical states are $\eta_{N}$ (with squared mass $m_{P}^{2}$) and
$\eta_{S}$ (with squared mass $m_{P}^{2}+\delta_{S}$). Last, the octet-singlet
splitting is parametrized by%

\begin{equation}
\mathcal{L}_{P,0}=-\alpha_{P}\eta_{0}^{2}=-\alpha_{P}\left(  \sqrt{2}\eta
_{N}+\eta_{S}\right)  ^{2}\text{ ,}\label{singlet}%
\end{equation}
where $\alpha_{P}=\alpha_{P,gg}+\alpha_{P,A}.$ Here, $\alpha_{P,gg}$ describes
processes with two intermediate transverse gluons ($\bar{n}n\rightarrow\bar
{n}n$, $\bar{n}n\rightarrow\bar{s}s,$ etc. ). This is a small perturbation.
The parameter $\alpha_{P,A}$ represents an effective contribution of the axial
anomaly; Eq. (\ref{singlet}) with $\alpha_{P}\simeq\alpha_{P,A}$ was also
obtained in\ e.g. Refs. \cite{rosenzweig,psg}. If $\mathcal{L}_{P,U(3)}%
+\mathcal{L}_{P,0}$ is considered, the physical states are $\eta_{8}$ (with
squared mass $m_{P}^{2}$) and $\eta_{0}$ (with squared mass $m_{P}^{2}%
+2\alpha_{P}$). Thus, $\mathcal{L}_{P,S}$ and $\mathcal{L}_{P,0}$ lead to
different basis, and the question is which one is dominant.

In the full case, one considers $\mathcal{L}_{U(3)}+\mathcal{L}_{S}%
+\mathcal{L}_{P,0}$. The pseudoscalar mixing angle $\theta_{P}$ can be
calculated by the previous expressions: $\theta_{P}=-\frac{1}{2}\arctan\left[
\frac{4\sqrt{2}\alpha_{P}}{2(m_{K}^{2}-m_{\pi}^{2}-\alpha_{P})}\right]  $.
Numerically, $\theta_{P}$ varies between $-40^{\circ}$ and $-45^{\circ}%
$~\cite{feldmann,escribano,dick,kloe}. The mixing is rather large and the
states are closer to octet and singlet ones, but the effect of the $s$-quark
is also important. Note, in the limit $\alpha_{P}=0$ one gets $\theta_{P}=0$
(purely strange and nonstrange states). On the contrary, in the limit
$m_{K}^{2}-m_{\pi}^{2}=0$ ($\delta_{P,S}=0)$ one has $\theta_{P}=\frac{1}%
{2}\arctan\left[  2\sqrt{2}\right]  =35.3^{\circ}$, i.e. octet and singlet
states, see Eq. (\ref{mixing}).

In the recent work of Ref. \cite{anomaly}, it was shown that the
(pseudo)scalar multiplet is \textit{heterochiral}. Namely, it is described by
a matrix $\Phi$ (see \cite{dick}) which under chiral transformation changes as
$\Phi\rightarrow e^{-i\alpha}U_{L}\Phi U_{R}^{\dagger}$ (the parameter
$\alpha$ refers to $U(1)_{A}$). The Lagrangian $\mathcal{L}_{\Phi
}^{\text{anomaly}}=-a_{\mathrm{A}}^{(3)}[\mathrm{det}(\Phi)-\mathrm{det}%
(\Phi^{\dagger})]^{2}\,$preserves chiral symmetry but breaks $U(1)_{A}$ (this
is a consequence of the determinant, see also Ref. \cite{salamanca}). This
Lagrangian term reduces to Eq. (\ref{singlet}) when condensation is considered
and quadratic mass terms are isolated. In conclusion, the heterochiral
(pseudo)scalar nonet can easily explain the emergence of an anomalous term
affecting $\eta$ and $\eta^{\prime}.$

\section{Vector (and tensor) mesons}

Next, we consider the isoscalar vector states $\omega(782)$ and $\phi(1020)$.
Just as before, one introduces the nonstrange-strange basis $\omega_{N}%
=\sqrt{1/2}(\bar{u}u+\bar{d}d)$, $\omega_{S}=\bar{s}s$ and the octet-singlet
basis $\omega_{8}=\sqrt{1/6}(\bar{u}u+\bar{d}d-2\bar{s}s)$, $\omega_{0}%
=\sqrt{1/3}(\bar{u}u+\bar{d}d+\bar{s}s)$, for which Eq. (\ref{mixing}) holds
(upon, of course, renaming the fields). Also here, we consider three
Lagrangians:
\begin{equation}
\mathcal{L}_{V,U(3)}=-\frac{m_{V}^{2}}{2}\left(  \omega_{N}^{\mu2}+\omega
_{S}^{\mu2}\right)  \text{ , }\mathcal{L}_{V,S}=-\frac{\delta_{V,S}}{2}%
\omega_{S}^{\mu2}\text{ , }\mathcal{L}_{V,0}=-\alpha_{V}\omega_{0}^{2}\text{
.}%
\end{equation}
There is an important difference in the last term. For vector states, the
constant $\alpha_{V}=\alpha_{V,ggg}$ (three-gluon mixing processes, typically
small): there is \textit{no} contribution from the axial anomaly,
$\alpha_{V,A}=0$. As a consequence, $\omega(782)$ is basically nonstrange and
$\phi(1020)$ strange (mixing angle $\theta_{V}=-3^{\circ}$, equation analogous
to Eq. (\ref{mixing}) \cite{pdg}). Similarly, for their axial-vector chiral
partners it holds that: $f_{1}(1285)$ is almost purely nonstrange and
$f_{1}(1420)$ purely strange \cite{florianlisa}.

In Ref. \cite{anomaly} it was discussed why $\alpha_{V,A}=0.$ This is due to
the fact that the corresponding chiral multiplets of vector ($V_{\mu}$) and
axial-vector ($A_{\mu}$) states are \textit{homochiral}. Namely, they enter
into the right(left)-handed $R_{\mu}=V_{\mu}-A_{\mu}\,$and $L_{\mu}=V_{\mu
}+A_{\mu},$ which under chiral symmetry transforms as $L_{\mu}\longrightarrow
U_{\mathrm{L}}\,R_{\mu}\,U_{\mathrm{L}}^{\dagger}\,,$ $R_{\mu}\longrightarrow
U_{\mathrm{R}}\,R_{\mu}\,U_{\mathrm{R}}^{\dagger}\,$ (in both cases, either
only $U_{\mathrm{L}}$ or $U_{\mathrm{R}}\,$appears, but no mixed terms). There
is no term involving the determinant.

A similar analysis applies to the ground-state tensor mesons, which are also
part of an heterochiral multiplet: $f_{2}(1270)$ is almost purely nonstrange
and $f_{2}^{\prime}(1525)$ strange, in agreement with the phenomenology
\cite{tensor}.

\section{Pseudotensor mesons}

In the end, we consider the pseudotensor sector. We start from $\eta
_{2,N}=\sqrt{1/2}(\bar{u}u+\bar{d}d)$, $\eta_{S}=\bar{s}s$ and $\eta
_{2,8}=\sqrt{1/6}(\bar{u}u+\bar{d}d-2\bar{s}s)$, $\eta_{2,0}=\sqrt{1/3}%
(\bar{u}u+\bar{d}d+\bar{s}s)$. The Lagrangian terms read%

\begin{equation}
\mathcal{L}_{PT,U(3)}=-\frac{m_{PT}^{2}}{2}\left(  \eta_{2,N}^{\mu\nu,2}%
+\eta_{2,S}^{\mu\nu,2}\right)  \text{ , }\mathcal{L}_{PT,S}=-\frac{\delta_{S}%
}{2}\eta_{2,S}^{\mu\nu2}\text{ , }\mathcal{L}_{PT,0}=-\alpha_{PT}\eta
_{2,0}^{\mu\nu,2}%
\end{equation}
Here, $\alpha_{PT}=\alpha_{PT,gg}+\alpha_{PT,A},$ and the latter quantity is
expected to be sizable, hence the anomaly is potentially large. This is due to
the fact that the corresponding chiral multiplet $\Phi_{\mu\nu}$ is
heterochiral, just as for pseudoscalar mesons. In fact, under chiral
transformations it transforms as $\Phi_{\mu\nu}\rightarrow\mathrm{e}%
^{-\mathrm{i}\alpha}\,U_{\mathrm{L}}\,\Phi_{\mu\nu}\,U_{\mathrm{R}}^{\dagger}$
\cite{anomaly}. The corresponding Lagrangian term $\mathcal{L}_{\Phi_{\mu\nu}%
}^{\text{anomaly}}\propto(\varepsilon^{ijk}\varepsilon^{i^{\prime}j^{\prime
}k^{\prime}}\Phi^{ii^{\prime}}\Phi^{jj^{\prime}}\Phi_{\mu\nu}^{kk^{\prime}%
}-h.c.)^{2}$ is chirally symmetric but breaks $U(1)_{A}$ (it is an extension
of the determinant) and reduces to $\mathcal{L}_{PT,0}$ when the condensation
of $\Phi$ is considered. The physical fields $\eta_{2}(1645)$ octet, $\eta
_{2}(1870)$ are:
\begin{equation}%
\begin{pmatrix}
\eta_{2}(1645)\\
\eta_{2}(1870)
\end{pmatrix}
=%
\begin{pmatrix}
\cos\theta_{PT} & \sin\theta_{PT}\\
-\sin\theta_{PT} & \cos\theta_{PT}%
\end{pmatrix}%
\begin{pmatrix}
\eta_{2,N}\\
\eta_{2,S}%
\end{pmatrix}
\text{ ,\thinspace}%
\end{equation}
with $\theta_{PT}\simeq-\frac{1}{2}\arctan\left[  \frac{4\sqrt{2}\alpha_{PT}%
}{2(m_{K_{2}(1770)}^{2}-m_{\pi_{2}(1660)}^{2}-\alpha_{PT})}\right]  $.
According to the phenomenological study of Ref. \cite{adrian}, $\theta
_{PT}\simeq-42^{\circ}$: a surprisingly large and negative mixing (similar to
the pseudoscalar sector) is realized, a fact that can be nicely explained by
the axial anomaly being important in this (heterochiral) sector.

\section{Conclusions}

We have studied the role of the axial anomaly for light mesons. For the
so-called \textquotedblleft heterochiral\textquotedblright\ multiplets
\cite{anomaly} (pseudoscalar and pseudotensor states), a large
strange-nonstrange mixing is expected (a known fact for $\eta$ and
$\eta^{\prime}$, some experimental evidence exists for pseudotensor mesons
\cite{adrian}). On the contrary, (axial-)vector and tensor mesons are
\textquotedblleft homochiral\textquotedblright\ and the anomaly does not
affect the mixing: the isoscalar states are (almost) nonstrange and strange,
respectively. Ongoing experimental activity at the JLab (e.g. Ref.
\cite{gluex}) can help to shed light on resonances between 1-2 GeV and hence
on the role of the axial anomaly.

As recently shown, the axial anomaly can also be relevant in the baryonic
sector. In particular, it can explain the large decay $N(1535)\rightarrow
N\eta$ \cite{olbrich} and contribute to pion-nucleon scattering \cite{phillip}%
. Moreover, the enigmatic pseudoscalar glueball \cite{psg} is also related to
the axial anomaly and can be studied in the future.

\bigskip

\textbf{Acknowledgments}: the author thanks R. Pisarski and A. Koenigstein for
cooperation. Financial support from the Polish National Science Centre NCN
through the OPUS project no.~2015/17/B/ST2/01625 is acknowledged.


\begin{thebibliography}{99}                                                                                               %


\bibitem {thooft}G.~'t Hooft,
%``How Instantons Solve the U(1) Problem,''
Phys.\ Rept.\ \textbf{142}  357 (1986). D.~J.~Gross, S.~B.~Treiman and
F.~Wilczek,
%``Light Quark Masses and Isospin Violation,''
Phys.\ Rev.\ D \textbf{19} 2188 (1979) . S.~L.~Adler,
%``Axial vector vertex in spinor electrodynamics,''
Phys.\ Rev.\ \textbf{177} (1969) 2426. J.~S.~Bell and R.~Jackiw,
%``A PCAC puzzle: pi0 --> gamma gamma in the sigma model,''
Nuovo Cim.\ A \textbf{60} 47 (1969).

\bibitem {feldmann}%\cite{Schechter:1992iz}
%\bibitem{Schechter:1992iz}
J.~Schechter, A.~Subbaraman and H.~Weigel,
%``Effective hadron dynamics: From meson masses to the proton spin puzzle,''
Phys.\ Rev.\ D {\bf 48} 339 (1993).
%doi:10.1103/PhysRevD.48.339
%[hep-ph/9211239].
%%CITATION = doi:10.1103/PhysRevD.48.339;%%
T.~Feldmann, P.~Kroll and
B.~Stech,
%``Mixing and decay constants of pseudoscalar mesons,''
Phys.\ Rev.\ D \textbf{58} (1998) 114006. S.~D.~Bass and A.~W.~Thomas,
%``eta bound states in nuclei: A Probe of flavor-singlet dynamics,''
Phys.\ Lett.\ B \textbf{634} (2006) 368. G.~Amelino-Camelia \textit{et al.},
%``Physics with the KLOE-2 experiment at the upgraded DA$\phi$NE,''
Eur.\ Phys.\ J.\ C \textbf{68} 619 (2010). B.~Borasoy and R.~Nissler,
%``Hadronic eta and eta-prime decays,''
Eur.\ Phys.\ J.\ A \textbf{26} 383 (2005).

\bibitem {escribano}
%\cite{Escribano:2005qq}
R.~Escribano and J.~M.~Frere,
%``Study of the eta - eta-prime system in the two mixing angle scheme,''
JHEP \textbf{0506} (2005) 029.
%doi:10.1088/1126-6708/2005/06/029
%[hep-ph/0501072].
%%CITATION = doi:10.1088/1126-6708/2005/06/029;%%
%\cite{Bramon:1997va}
%\bibitem{Bramon:1997va}
A.~Bramon, R.~Escribano and M.~D.~Scadron,
%``The eta - eta-prime mixing angle revisited,''
Eur.\ Phys.\ J.\ C \textbf{7} 271 (1999).
%doi:10.1007/s100529801009
%[hep-ph/9711229].
%%CITATION = doi:10.1007/s100529801009;%%
%\cite{Escribano:2007cd}
R.~Escribano and J.~Nadal,
%``On the gluon content of the eta and eta-prime mesons,''
JHEP \textbf{0705}  006 (2007).
%doi:10.1088/1126-6708/2007/05/006
%[hep-ph/0703187].
%%CITATION = doi:10.1088/1126-6708/2007/05/006;%%


\bibitem {anomaly}
%\cite{Giacosa:2017pos}
%\bibitem{Giacosa:2017pos}
%\cite{Giacosa:2017pos}
%\bibitem{Giacosa:2017pos}
F.~Giacosa, A.~Koenigstein and R.~D.~Pisarski,
%``How the axial anomaly controls flavor mixing among mesons,''
Phys.\ Rev.\ D \textbf{97}  no.9  091901 (2018).
%doi:10.1103/PhysRevD.97.091901 [arXiv:1709.07454 [hep-ph]]
%%CITATION = doi:10.1103/PhysRevD.97.091901;%%


\bibitem {salamanca}
%\cite{Giacosa:2017ojs}
%\bibitem{Giacosa:2017ojs}
F.~Giacosa,
%``Revisiting the axial anomaly for light mesons and baryons,''
PoS Hadron \textbf{2017} 045 (2018).
%doi:10.22323/1.310.0045 [arXiv:1712.04664 [hep-ph]].


\bibitem {adrian}A.~Koenigstein and F.~Giacosa,
%``Phenomenology of pseudotensor mesons and the pseudotensor glueball,''
Eur.\ Phys.\ J.\ A \textbf{52}  no.12 (2016).

\bibitem {dick}D.~Parganlija, P.~Kovacs, G.~Wolf, F.~Giacosa and
D.~H.~Rischke,
%``Meson vacuum phenomenology in a three-flavor linear sigma model with (axial-)vector mesons,''
Phys.\ Rev.\ D \textbf{87} no.1 014011 (2013) . S.~Janowski, F.~Giacosa and
D.~H.~Rischke,
%``Is f0(1710) a glueball?,''
Phys.\ Rev.\ D \textbf{90}  no.11, 114005 (2014).

\bibitem {rosenzweig}C.~Rosenzweig, J.~Schechter and C.~G.~Trahern,
%``Is the Effective Lagrangian for QCD a Sigma Model?,''
Phys.\ Rev.\ D \textbf{21}  3388 (1980). A.~H.~Fariborz, R.~Jora and
J.~Schechter,
%``Toy model for two chiral nonets,''
Phys.\ Rev.\ D \textbf{72} 034001 (2005).

\bibitem {psg}W.~I.~Eshraim, S.~Janowski, F.~Giacosa and D.~H.~Rischke,
%``Decay of the pseudoscalar glueball into scalar and pseudoscalar mesons,''
Phys.\ Rev.\ D \textbf{87} no.5, 054036 (2013). W.~I.~Eshraim, S.~Janowski,
A.~Peters, K.~Neuschwander and F.~Giacosa,
%``Interaction of the pseudoscalar glueball with (pseudo)scalar mesons and nucleons,''
Acta Phys.\ Polon.\ Supp.\ \textbf{5}, 1101 (2012).

\bibitem {kloe}
%\cite{AmelinoCamelia:2010me}
%\bibitem{AmelinoCamelia:2010me}
G.~Amelino-Camelia \textit{et al.},
%``Physics with the KLOE-2 experiment at the upgraded DA$\phi$NE,''
Eur.\ Phys.\ J.\ C \textbf{68} 619 (2010).
%doi:10.1140/epjc/s10052-010-1351-1
%[arXiv:1003.3868 [hep-ex]].
%%CITATION = doi:10.1140/epjc/s10052-010-1351-1;%%


\bibitem {pdg}M. Tanabashi \textit{et al.} (Particle Data Group), Phys. Rev. D
98, 030001 (2018).

\bibitem {florianlisa}
%\cite{Divotgey:2013jba}
%\bibitem{Divotgey:2013jba}
F.~Divotgey, L.~Olbrich and F.~Giacosa,
%``Phenomenology of axial-vector and pseudovector mesons: decays and mixing in the kaonic sector,''
Eur.\ Phys.\ J.\ A \textbf{49} 135 (2013).
%doi:10.1140/epja/i2013-13135-3
%[arXiv:1306.1193 [hep-ph]].
%%CITATION = doi:10.1140/epja/i2013-13135-3;%%


\bibitem {tensor}
%\cite{Giacosa:2005bw}
%\bibitem{Giacosa:2005bw}
F.~Giacosa, T.~Gutsche, V.~E.~Lyubovitskij and A.~Faessler,
%``Decays of tensor mesons and the tensor glueball in an effective field approach,''
Phys.\ Rev.\ D \textbf{72} 114021 (2005).
%doi:10.1103/PhysRevD.72.114021
%[hep-ph/0511171].
%%CITATION = doi:10.1103/PhysRevD.72.114021;%%
L.~Burakovsky and J.~T.~Goldman,
%``Towards resolution of the enigmas of P wave meson spectroscopy,''
Phys.\ Rev.\ D \textbf{57} (1998) 2879. V.~Cirigliano, G.~Ecker, H.~Neufeld
and A.~Pich,
%``Meson resonances, large N(c) and chiral symmetry,''
JHEP \textbf{0306}  012 (2003).

\bibitem {gluex}H.~Al Ghoul \textit{et al.} [GlueX Collaboration],
%``First Results from The GlueX Experiment,''
AIP Conf.\ Proc.\ \textbf{1735}  020001 (2016). A.~Rizzo [CLAS
Collaboration],
%``The meson spectroscopy program with CLAS12 at Jefferson Laboratory,''
PoS CD \textbf{15}  060 (2016).

\bibitem {olbrich}
%\cite{Olbrich:2017fsd}
%\bibitem{Olbrich:2017fsd}
L.~Olbrich, M.~Z\'{e}t\'{e}nyi, F.~Giacosa and D.~H.~Rischke,
%``Influence of the axial anomaly on the decay $N(1535) \rightarrow N\eta $,''
Phys.\ Rev.\ D \textbf{97}  no.1, 014007 (2018).
%doi:10.1103/PhysRevD.97.014007 [arXiv:1708.01061 [hep-ph]].
%%CITATION = doi:10.1103/PhysRevD.97.014007;%%


\bibitem {phillip}
%\cite{Lakaschus:2018rki}
%\bibitem{Lakaschus:2018rki}
P.~Lakaschus, J.~L.~P.~Mauldin, F.~Giacosa and D.~H.~Rischke,
%``Role of a four-quark and a glueball state in pion-pion and pion-nucleon scattering,''
arXiv:1807.03735 [hep-ph].
%%CITATION = ARXIV:1807.03735;%%

\end{thebibliography}
\end{document}